\documentclass[11pt]{article}

\usepackage{cite}
\usepackage{amsmath}
\usepackage{amsfonts}
\usepackage{bm}

\def\lfrac#1#2{{}^{#1\!}\kern-.0em/_{#2}}
\def\buildrel#1\under#2{\mathrel{\mathop{\kern0pt #2}\limits_{#1}}}
\def\Res#1{{\buildrel {\scriptstyle #1} \under {\textstyle \rm Res}}\,}

\begin{document}

\vspace{2cm}

\noindent
{\sf CERN-TH/2001-185}\\[3ex]
\vspace{0.8cm}

\begin{center}
{\Large \sf QED Effective Action Revisited}
\end{center}

\vspace{0.05cm}

\begin{center}
U.~D.~Jentschura$^{a)}$, H.~Gies$^{b)}$, S.~R.~Valluri$^{c)}$,
D.~R.~Lamm$^{d)}$, and E.~J.~Weniger$^{e)}$
\end{center}

\vspace{0.05cm}

\begin{center}
$^{a)}${\it Institute of Theoretical Physics,}\\
{\it Dresden University of Technology, 01062 Dresden,
Germany}\\[1ex]
$^{b)}${\it CERN (Theory Division), CH-1211 Geneva 23, Switzerland}\\[1ex]
$^{c)}${\it Department of Applied Mathematics, Physics and Astronomy,}\\
{\it University of Western Ontario, London, N6A 3K7, Canada}\\[1ex]
$^{d)}${\it Georgia Institute of Technology (GTRI/EOEML),}\\
{\it Atlanta, Georgia GA 30332--0834, USA}\\[1ex]
$^{e)}${\it Institute of Physical and Theoretical Chemistry,}\\
{\it University of Regensburg, D-93040 Regensburg, Germany}\\[1ex]
\end{center}

\vspace{0.6cm}

\begin{center}
\begin{minipage}{10.5cm}
{\underline{Abstract}}
The derivation of a convergent series representation for the quantum
electrodynamic effective action obtained by two of us (S.R.V.~and 
D.R.L.) in [Can. J. Phys. {\bf
71}, 389 (1993)] is reexamined. We present more details of our original
derivation. Moreover, we discuss the relation of the electric-magnetic
duality to the integral representation for the effective action, and we
consider the application of nonlinear convergence acceleration
techniques which permit the efficient and reliable numerical evaluation
of the quantum correction to the Maxwell Lagrangian.
\end{minipage}
\end{center}

\vspace{0.6cm}

Asymptotic problems and properties;\\
Quantum Electrodynamics -- Specific Calculations\\
\vfill
\begin{center}
\begin{minipage}{14cm}
\begin{center}
\hrule
{\bf \scriptsize
\noindent electronic mail:
  jentschura@physik.tu-dresden.de, holger.gies@cern.ch,\\[-1.0ex]
  darrell.lamm@gtri.gatech.edu, valluri@uwo.ca,
  joachim.weniger@chemie.uni-regensburg.de.}
\end{center}
\end{minipage}
\end{center}

\newpage

%
%
\section{Introduction}
\label{Introduction}

Maxwell's equations receive corrections from virtual excitations
of the charged quantum fields (notably electrons and positrons).
This leads to interesting effects~\cite{DiGi2000}:
light-by-light scattering, photon splitting, modification
of the speed of light in the presence of strong electromagnetic fields,
and -- last, but not least -- pair production. 

When the heavy degrees of freedom are integrated out (in this case, 
the ``heavy particles'' are the electrons and positrons), an effective
theory results. The corrections can be described by an effective interaction,
the so-called quantum electrodynamic (QED) effective Lagrangian.
The dominant effect for electromagnetic fields that    
vary slowly with respect to the Compton wavelength 
(frequencies $\omega \ll 2\,m\,c^2/\hbar$) is described by
the one--loop quantum electrodynamic effective
(so-called ``Heisenberg--Euler'') Lagrangian which is known to 
all orders in the electromagnetic 
field~\cite{HeEu1936,We1936,Sc1951,BBBB1970,DiRe1985,DiGi2000}. 
 
The Heisenberg--Euler Lagrangian $\Delta {\cal L}$, which
constitutes a quantum correction to the 
Maxwell Lagrangian, is usually expressed as a one-dimensional 
proper-time integral [see e.g. Eq.~(3.43) in~\cite{DiGi2000},
the notation is clarified in Sec.~\ref{RepSpecial} below]:
\begin{equation}
\label{EffAct}
\Delta {\mathcal L} = - \frac{e^2}{8 \pi^2} \,
\lim_{\epsilon,\eta \to 0^{+}} \,
\int_{{\mathrm i}\,\eta}^{\infty + {\mathrm i}\,\eta} \,
\frac{{\rm d} s}{s} \,
{\rm e}^{- (m^2 - {\rm i} \epsilon)\, s}\,
\biggl[  \, a b \, \coth(e a s) \, \cot(e b s) \, -
\frac{a^2 - b^2}{3} - \frac{1}{(e s)^2} \biggr] \,.
\end{equation}
Because there are
no singularities in the first quadrant of the complex plane
and because Jordan's Lemma may be applied to the
integral, it is possible to exchange the lower and upper
limit of integration by $\eta$ and
$\eta + {\mathrm i}\,\infty$, respectively.

Although the proper-time integral (\ref{EffAct}) can be evaluated by
numerical quadrature, it is evident that a representation by a
convergent series expansion could have certain computational as well as
conceptual advantages. An expansion of (\ref{EffAct}) in terms of 
special functions has been given by
us in~\cite{VaLaMi1993,VaMiLa1994}. Here, we present a unified
series expansion which encompasses both the real and the imaginary part of
the Lagrangian (see Sec.~\ref{RepSpecial}). Also, we clarify
certain technical details concerning the derivation of our previous
results~\cite{VaLaMi1993,VaMiLa1994}.
In Sec.~\ref{EMDuality}, we discuss the 
``electric-magnetic duality'' which has recently     
drawn much attention~\cite{ChPa2001,ChPa2001prd}.   
In particular, we elucidate different kinds of dual invariance and 
pursue the question as to whether these invariances are realised by 
QED effective Lagrangians. Before we expand on these aspects, we
would like to provide some general discussion on the general 
relevance of studies related to the QED effective action~(\ref{EffAct}).

The real part of the Lagrangian (\ref{EffAct}) 
can be used to delineate dispersive phenomena such as photon propagation
in a magnetic field~(see~\cite{DiGi2000} and references therein), 
photon splitting~\cite{Ad1971,St1979,Ba1995,DuTh1992},
vacuum birefringence and second harmonic generation~\cite{VaBa1980},
and light scattering in a vacuum~\cite{DiGi2000}.   
These applications have a strong relevance to particle astrophysics.
The imaginary part of the Lagrangian (\ref{EffAct})
has been applied to absorption processes such as electron-positron pair
creation and dichroism (see e.g.~\cite{Sc1951,VaLaMi1993,
KlNi1964,Er1966,HeHe1997jpa,ScEtAl1998}). The
occurrence of strong fields in storage rings, pulsars, magnetars and
high-intensity lasers also motivates our
study. The advent of state-of-the-art laser beams and photon detectors may
provide a signature of ``QED's nonlinear light''. The optical second
harmonic generation (SHG) in vacuo is an
interesting second-order magneto-optical effect that occurs if the
spatial symmetry of the nonlinearity induced by the effective
action (\ref{EffAct}) is broken e.g.~by a strong static magnetic field
(in a more general context, higher-harmonic generation by vacuum 
effects was discussed in~\cite{BaVa1978,VaBa1980}). 
Note that the leading contribution to SHG in vacuo involves a fourth-order
effect in contrast to the much weaker 6th-order effect (hexagon
graph) which gives the leading contribution to photon splitting.
SHG and photon splitting in vacuo may be 
occurring close to astronomical objects such as white 
dwarfs, neutron stars and ``magnetars'' (see~\cite{Ba1995,HeHe1997}
and references therein) which have strong
magnetic fields up to $B \approx 10^{14}~{\mathrm G}$. 
In all cases, a detailed, realistic description of the 
experimental conditions and/or the involved astrophysical
objects, especially at extreme field strengths, 
requires techniques for the reliable numerical
evaluation of the QED effective action.

{\em A priori}, the construction of a series expansion for (\ref{EffAct})
constitutes a complete solution of the problem
from a theoretical point of view.
However, such an expansion does not necessarily provide all answers:
Many series expansions are
known which either converge extremely slowly or which do not converge at
all.  Moreover, in the case
of the Heisenberg--Euler
Lagrangian, there is the additional problem that the terms of the series
are represented by special
functions which are in most cases defined and computed via series
expansions.  Again, convergence
problems are more likely the rule than the exception. 
Here, we are concerned with the solution of the principal numerical
difficulty associated with the the slow overall convergence of the 
series expansion derived in~\cite{VaLaMi1993}, whose
terms are nonalternating in sign. 
The modern theory of nonlinear sequence transformations which begins
with Wynn's epsilon algorithm~\cite{Wy1956a,Wy1956b}
was developed when the first computers became generally available. 
Pad\'{e} approximants, however, would not be powerful
enough to sum our series for the Heisenberg--Euler
Lagrangian derived in~\cite{VaLaMi1993}.  

From a mathematical point of view, there is a distant analogy between
the expansion of (\ref{EffAct}) in terms of special functions and the
(exact) expansion of certain quantum electrodynamic bound-state
effects into partial
waves~\cite{Mo1974a,Mo1974b,JeMoSo1999,JeMoSo2001}. However, the
mathematical entities involved in the present decomposition possess a
far less involved mathematical structure, and it is difficult to
associate a meaningful physical interpretation to each term in the
special function representation of~(\ref{EffAct}). As is evident from
the discussion in Sec.~\ref{ImpIdentity}, the terms of the convergent
series representation may be interpreted as being generated by a
``partial-fraction decomposition'' in distant analogy to the
``partial-wave decomposition'' in bound-state calculations.  We point
out in~Sec.~\ref{AccConv} that the convergence of the special function
representation can be accelerated by the same technique -- the
so-called ``Combined Nonlinear--Condensation Transformation''
(CNCT)~\cite{JeMoSoWe1999} -- which was used successfully for the
acceleration of the convergence of partial-wave decompositions in
quantum electrodynamic bound-state
calculations~\cite{JeMoSo1999,JeMoSo2001}.

At the same time, we would like to mention that 
the integral (\ref{EffAct}), when expanded in powers of the 
electric and magnetic field strengths, represents a divergent
series. The resulting divergent series can be used as a
``model laboratory'' for resummation 
methods~\cite{DuHa1999,CvYu1999,JeWeSo2000,JeBeWeSo2000},
and related investigations~\cite{JeWeSo2000,JeBeWeSo2000,JeSo2001,Je2001pra} 
have led to the development of asymptotically improved resummation algorithms
which have recently received interesting 
applications~\cite{CvLe2001,CvKo2001,CvDiLeSc2001}
in the highly accurate determination of the 
strong coupling constant at the $Z$ pole and
other investigations on nonperturbative effects
in gauge theories~\cite{ChElSt1999a,ChElSt1999b,ChElSt2000,ElChSt2000}.
These investigations are related to the 
fundamental question of how to ``make sense'' of the fact that many
perturbation series encountered in physics are divergent,
and are {\em not} meant to provide efficient means for 
numerical evaluation of the integral~(\ref{EffAct}).

%
%
\section{Representation of the QED Effective Action by\\
Special Functions}
\label{RepSpecial}

The QED effective Lagrangian can be expressed as a function
of the Lorentz invariants ${\mathcal F}$ and 
${\mathcal G}$ which are given by
\begin{eqnarray} 
\label{DefF}
{\mathcal F} & = 
\lfrac{1}{4} \, F_{\mu\nu} \, F^{\mu\nu} =
\lfrac{1}{2} \, 
\left(\bm{B}^2 - \bm{E}^2\right) = &
\lfrac{1}{2} \, \left(a^2 - b^2\right)\,, \\[2ex]
\label{DefG}
{\mathcal G} & = 
\lfrac{1}{4} \, F_{\mu\nu} \, (*F)^{\mu\nu} \;\; =
\;\; -\bm{E}\cdot\bm{B} \;\; = & \pm a b\,,
\end{eqnarray}
where $\bm{E}$ and $\bm{B}$ are the 
electric and magnetic field strengths, 
$F_{\mu\nu}$ is the field strength tensor,
and $(*F)^{\mu\nu}$ denotes the 
dual field strength tensor
$(*F)^{\mu\nu} = (1/2) \, \epsilon^{\mu\nu\rho\sigma}\,F_{\rho\sigma}$.
By $a$ and $b$ we denote the {\em secular invariants},
\begin{eqnarray}
\label{Defab}
a &=& \sqrt{\sqrt{{\mathcal F}^2 +
{\mathcal G}^2} + {\mathcal F}}\,,\nonumber\\
b &=& \sqrt{\sqrt{{\mathcal F}^2 +
{\mathcal G}^2} - {\mathcal F}}\,.
\end{eqnarray}
These Lorentz invariants are referred to as secular invariants
because they emerge naturally as eigenvalues of the field strength
tensor; these eigenvalues are conserved under proper 
Lorentz transformations of the field strength tensor.
There are connections between the different representations~\cite{DiGi2000}:
If the relativistic invariant ${\mathcal G}$ is positive,
then it is possible to transform to a Lorentz frame in which $\bm{E}$ and
$\bm{B}$ are {\em antiparallel}. In the case ${\mathcal G} < 0$,
it is possible to choose a Lorentz frame in which  $\bm{E}$ and
$\bm{B}$ are {\em parallel}. Irrespective of the sign
of ${\mathcal G}$ we have in the specified frame
\begin{equation}
a = |\bm{B}| \quad \mbox{and} \quad b = |\bm{E}| 
\quad \mbox{if and only if} \quad 
\mbox{$\bm{B}$ is (anti-)parallel to $\bm{E}$}\,.
\end{equation}
In any case, because $a$ and $b$ are positive definite, 
we have
\begin{equation}
a \, b = |\bm{E} \cdot \bm{B}| > 0 \quad \mbox{for} \quad 
\mbox{any Lorentz frame and ${\mathcal G} \neq 0$}\,,
\end{equation}
which clarifies the sign ambiguity in (\ref{DefG}).
We give in (\ref{DefF}) and (\ref{DefG}) seemingly redundant definitions,
but it will soon become apparent that each of the alternative
``points of view'' has its applications.
The Maxwell Lagrangian is given by
\begin{equation}
{\cal L}_{\text{cl}} = - {\mathcal F} = 
-\lfrac{1}{4} \, F_{\mu\nu} \,
F^{\mu\nu} = 
\lfrac{1}{2}\,
\left(\bm{E}^2 - \bm{B}^2\right) =
\lfrac{1}{2} \, \left(b^2 - a^2\right)\,.
\end{equation}
As it is obvious from Eq.~(\ref{EffAct}),
the correction $\Delta {\cal L}$ to the
Maxwell Lagrangian is conveniently written 
in terms of the secular invariants $a$ and $b$.

The effective action has a dispersive (real) part and
an imaginary part which is associated with pair production,
\begin{equation}
\Delta {\mathcal L} = {\rm Re}\,{\Delta {\cal L}}
+ {\mathrm i}\, {\rm Im}\,{\Delta {\cal L}}\,.
\end{equation}
In Eqs.~(2)--(6)
of~\cite{VaLaMi1993} we showed that the real part of 
(\ref{EffAct}) can be expressed as
\label{resre}
\begin{eqnarray}
\label{RepSpecReal}
\lefteqn{{\rm Re}\,{\Delta {\cal L}} \!
= \! - \frac{e^2}{4\,\pi^3} \, a\, b\,
\sum_{n=1}^{\infty} \, \bigl[ a_n + d_n \bigr] \, ,} \nonumber\\[2ex]
a_n & \!\! = & \frac{\coth (n \pi b/a)}{n} \, \biggl\{
{\rm Ci} \left( \frac{n \pi m^2}{e a} \right) \,
\cos \left( \frac{n \pi m^2}{e a} \right) \, + \,
{\rm si} \left( \frac{n \pi m^2}{e a} \right) \,
\sin \left( \frac{n \pi m^2}{e a} \right) \biggr\} \, ,
\nonumber\\[2ex]
d_n & \!\! = & \!\! - \frac{\coth (n \pi a/b)}{2 n} \, \biggl\{
\exp \left( \frac{n \pi m^2}{e b} \right) \,
{\rm Ei} \left( - \frac{n \pi m^2}{e b} \right)
\, + \, \exp \left( - \frac{n \pi m^2}{e b} \right) \,
{\rm Ei} \left( \frac{n \pi m^2}{e b} \right) \biggr\} \,.
\end{eqnarray}
We also derived the following representation 
for the imaginary part [see Eq.~(18) of~\cite{VaLaMi1993}]:
\begin{equation}
\label{RepSpecIm}
{\rm Im}\,{\Delta {\cal L}} =
\frac{e^2 a b}{8\,\pi^2} \sum_{n=1}^{\infty} \frac{1}{n}
\coth\left( \frac{n\pi a}{b}\right)
\exp\left(- \frac{n\pi m^2}{e\,b}\right) \, .
\end{equation}
These results have recently been 
confirmed in~\cite{ChPa2001}.
For the special functions,
we use the notation of Abramowitz and Stegun~\cite{AbSt1972}.
Here, one might wonder why the cosine and sine integrals 
appear in an ``asymmetric'' form (Ci and si instead
of ci and si) in the definitions
of $a_n$ and $d_n$ in (\ref{RepSpecReal}). The reason is that
the commonly accepted definitions for the cosine and sine
integrals are ``asymmetric'' in the following sense
[see Eqs.~(5.2.1),
(5.2.2), (5.2.5), (5.2.26), and (5.2.27) of~\cite{AbSt1972}]:
\begin{eqnarray}
\label{defs}
{\rm Ci}(z) &=& - \int_z^{\infty} {\rm d}t \, \frac{\cos(t)}{t} \,,
\nonumber\\
{\rm si}(z) &=& - \int_z^{\infty} {\rm d}t \, \frac{\sin(t)}{t} =
{\rm Si}(z) - \frac{\pi}{2}\,,
\nonumber\\
{\rm Si}(z) &=& \int_0^{z} {\rm d}t \, \frac{\sin(t)}{t}\,.
\end{eqnarray}
From these formulas, it is evident that ``symmetric'' integrals
with lower limit $z$ and upper limit $\infty$ require 
the ``asymmetric'' occurrence of Ci and si.

Because the imaginary part ${\rm Im}\,{\Delta {\cal L}}$
is generated exclusively by analytic continuation of
one of the exponential integrals -- specifically, the 
term 
\[
{\rm Ei} \left( - \frac{n \pi m^2}{e b} \right)
\]
in the definition of $d_n$ --,  it is obvious 
how to write down a unified representation for both 
the real and the imaginary part.
We therefore present here the 
unified representation for both the real
and the imaginary part we obtained in Eq.~(8) of
version 2 of our preprint~\cite{LaVaJeWe2000}:
\begin{eqnarray}
\label{UniSpecial}
\lefteqn{{\Delta {\cal L}} = \lim_{\epsilon \to 0^{+}}
- \frac{e^2}{4\,\pi^3} \, a\, b\,
\sum_{n=1}^{\infty} \, [b_n \, + \, c_n]\,, }\nonumber\\[3ex]
b_n & = & \!\!\!\! - \frac{\coth\left(n\pi b/a\right)}{2 n} \,
\left\{ \exp\left( - {\rm i} \, \frac{n\pi m^2}{e\,a}\right) 
\Gamma\left( 0, - {\rm i} \, \frac{n\pi m^2}{e\,a}\right) +
\exp\left( {\rm i} \, \frac{n\pi m^2}{e\,a}\right)
\Gamma\left( 0, {\rm i} \, \frac{n\pi m^2}{e\,a}\right) \right\} \,,
\nonumber\\[3ex]
c_n & = &\!\!\!\!\! \frac{\coth\left(n\pi a/b\right)}{2 n} \,
\left\{ \exp\left( \frac{n\pi m^2}{e\,b}\right)
{\Gamma} \left(0, \frac{n\pi m^2}{e\,b}\right) + \!
\exp\left( - \frac{n\pi m^2}{e\,b}\right) 
\Gamma\left(0, -\frac{n\pi m^2}{e\,b} +
  {\rm i}\,\epsilon \right) \right\} \! .
\end{eqnarray}
It becomes obvious from this representation that 
the effective action has branch cuts along the positive
and negative $b$ axis as well as the positive and negative  
imaginary $a$ axis.
Here, we make extensive use of the 
incomplete Gamma function defined as [see Eq.~(6.5.3) of~\cite{AbSt1972}]
\begin{equation}
\label{incgam}
\Gamma(a, z) = \int_z^{\infty} {\rm d}t\, {\rm e}^{-t} \, t^{a-1} \,.
\end{equation}
For $a = 0$, the quantity $\Gamma(0, z)$ as a function of $z$
has a branch cut along the negative
real $z$ axis, and we assume 
\begin{equation}
\lim_{\epsilon \to 0^{+}}
{\rm Im} \, \Gamma(0, -x + {\rm i} \, \epsilon) \; = \;
- \pi \, , \qquad x > 0
\end{equation}
which follows from the relationships~\cite[Eq.~(5.1.45)]{AbSt1972}
$E_1 (z) = \Gamma (0, z)$ and~\cite[Eq.~(5.1.7)]{AbSt1972} $E_1
(-x \pm \mathrm{i}0) = - \mathrm{Ei} (x) \mp \mathrm{i} \pi$. 

A unified expansion in terms of special functions -- including infinitesimal 
imaginary parts -- has also been given in 
the final version of~\cite{ChPa2001}.
In this context it is perhaps worth pointing out that
it is inconsistent with standard notation to use the exponential
integral Ei for such a unified formula.
The exponential integral Ei is usually
defined only for real argument.
It is defined as a Cauchy principal-value
integral by Gradshtein and Ryzhik~\cite{GrRy1981}
[see Eqs.~(8.211.1) and~(8.211.2) {\em ibid.}] as well as
by Abramowitz and Stegun~\cite{AbSt1972} [see Eq.~(5.1.2)~{\em ibid.}],
and also predominantly in the mathematical literature;
see, for example, Olver~\cite{Ol1974} [see Eq.~(3.07) {\em ibid.}].
In contrast to the exponential integral Ei,
the incomplete Gamma function is defined in the entire complex 
plane with a cut along the negative real axis.

%
%
\section{An Important Mathematical Identity}
\label{ImpIdentity}

W.~J.~Mielniczuk \cite{Mi1982} has outlined 
a proof of the representation (\ref{RepSpecReal})
for the real part of the effective action.
However, his work suffered from a series of unfortunate typographical 
errors. Here, we provide details on the 
intermediate steps used in our 
calculation~\cite{VaLaMi1993}, and~we
give, in particular, a corrected version of identity (2.8) of~\cite{Mi1982}.
This corrected version was also used in obtaining the 
results in Eqs.~(\ref{RepSpecReal}) and 
(\ref{RepSpecIm}) above and in
Eqs.~(2) --- (6) and (18) in~\cite{VaLaMi1993}.

The corrected version of identity (2.8)
of~\cite{Mi1982} reads:
\begin{eqnarray}
\label{id}
\lefteqn{{\tilde x} \, {\tilde y} \, u^2 \, \coth({\tilde x} \, u) \,
\cot({\tilde y} \, u) - 1 -
\frac{1}{3} \, ({\tilde x}^2 - {\tilde y}^2)\,u^2 =}\nonumber\\
& & - \frac{2 \, {\tilde x}^3 \, {\tilde y} \, u^4}{\pi} \,
\sum_{k=1}^{\infty} \frac{1}{k} \,
\frac{1}{{\tilde x}^2 \, u^2 + k^2 \, \pi^2} \,
\coth\left( \frac{\tilde y}{\tilde x} \, k \, \pi \right) \nonumber\\
& & + \frac{2 \, {\tilde y}^3 \, {\tilde x} \, u^4}{\pi} \,
\sum_{k=1}^{\infty} \frac{1}{k} \,
\frac{1}{{\tilde y}^2 \, u^2 - k^2 \, \pi^2} \,
\coth\left( \frac{\tilde x}{\tilde y} \, k \, \pi \right) \, .
\end{eqnarray}
In Ramanujan's notebooks~\cite{Be1989}, this identity appears as
entry (19.3) on p.~271. A proof is given which 
is based on the repeated
application of the well-known identities [see {\em e.g.} Eq.~(19.2) 
of~\cite{Be1989} or Eq.~(2.4) of~\cite{VaLaMi1982}]
\begin{equation}
\label{cotexpansion}
\pi\,x\cot (\pi x) = 1 + 2 x^2
\, \sum_{n=1}^{\infty} 1/(x^2 - n^2)
\end{equation}
and 
\begin{equation}
\label{cothexpansion}
\pi\,y\coth (\pi y) = 1 + 2
y^2 \, \sum_{n=1}^{\infty} 1/(y^2 + n^2)\,,
\end{equation}
respectively, to each one
of the factors on the left-hand side of (\ref{id}), and a skillful
reformulation of the resulting double sum. 

Here, we will derive an alternative,
but fully equivalent
formulation of (\ref{id}):
\begin{eqnarray}
\label{id2}
\lefteqn{ab \coth (az) \cot (bz) \, - \, \frac{1}{z^2} \, - \,
\frac{a^2 - b^2}{3}} \nonumber \\
& = &  \frac{2abz^2}{\pi} \, \left\{ \sum_{k=1}^{\infty} \,
\frac{\coth (k \pi a / b)}{k \, [z^2 - k^2\pi^2/b^2]} \, - \,
\sum_{k=1}^{\infty} \,
\frac{\coth (k \pi b / a)}{k \, [z^2 + k^2\pi^2/a^2]} \right\} \, .
\end{eqnarray}
In our derivation of identity (\ref{id})
in~\cite{VaLaMi1993},
we used the so-called Partial
Fraction Theorem 4.4.5 of~\cite{MaHo1987}. It may be surprising that
this identity can be obtained by a straightforward 
application of this basic theorem to the left-hand side of
Eq.~(\ref{id}), especially in view of the fact that the derivation of
this result represented a considerable challenge to
S.~Ramanujan, as it is evident from remarks on p.~271
(top of page) of~\cite{Be1989}. This shows that sometimes interesting 
new results can be obtained by a
straightforward application of theorems occurring in standard
textbooks.

We quote here the Partial Fraction 
Theorem~\cite[Theorem 4.4.5 on pp.\ 337 - 338)]{MaHo1987}:
\begin{center}
\begin{minipage}{14.0cm}
{\it {\bf \sf Theorem 1}. 
Suppose that $g$ is meromorphic with simple poles at $a_1$, $a_2$,
$a_3$, \ldots with $0 \le \vert a_1 \vert \le \vert a_2 \vert \ldots$
and residues $b_k$ at $a_k$, but analytic at $0$. Suppose there is an
increasing sequence $R_1$, $R_2$, $R_3$, \ldots with
$\lim_{n \to \infty} R_n = \infty$ and simple closed curves
$\mathcal{C}_n$ satisfying
\begin{itemize}
\item[(i):] $\vert z \vert \ge R_n$ for all $z$ on $\mathcal{C}_n$,
\item[(ii):] There is a constant $S$ with {\rm length}$(\mathcal{C}_n) \le S
R_n$ for all $n \in \mathbb{N}$,
\item[(iii):] There is a constant $M$ with $\vert g(z) \vert \le M$ for
all $z$ on $\mathcal{C}_n$ and for all $n \in \mathbb{N}$.
\end{itemize}
Then,}
\begin{equation}
\label{f_MitLef}
g (z) \; = \; g (0) + \sum_{m=1}^{\infty} \,
\left\{ \frac{b_m}{z - a_m} + \frac{b_m}{a_m} \right\} \, .
\end{equation}
\end{minipage}
\end{center}

Here, it should be noted that normally the series on the right-hand side 
of (\ref{f_MitLef}) does not converge
absolutely, which implies that it must {\em not} be rearranged. In fact, the
series with the terms $b_m/(z-a_m)$ may even diverge (compare the remark
in the third paragraph on p.\ 539 of \cite{Osg1965}). Thus, the 
compensatory terms
$b_m/a_m$ are necessary to ensure convergence. 

Equation~(\ref{f_MitLef}) implies {\em eo ipso} -- just as the conditions of 
the partial fraction theorem -- that the function $g (z)$ which is expanded
into ``partial fractions'' must necessarily be analytic at zero argument.
In order to analyse the behaviour of the product $\coth (az) \cot
(bz)$ with $a, b \in \mathbb{R}$ as $z \to 0$, we use
\cite[Eq.\ (30:6.2)]{SpaOld1987}
\begin{equation}
\coth (z) \; = \; \frac{1}{z} \, \sum_{j=0}^{\infty} \,
\frac{4^j B_{2j}}{(2j)!} \, z^{2j} \, ,
\qquad \vert z \vert < \pi \, ,
\label{PowSerCOTH}
\end{equation}
and \cite[Eq.\ (34:6.2)]{SpaOld1987}
\begin{equation}
\cot (z) \; = \; \frac{1}{z} \, - \, \sum_{j=1}^{\infty} \,
\frac{4^j \vert B_{2j} \vert}{(2j)!} \, z^{2j-1} \, ,
\qquad \vert z \vert < \pi \, .
\label{PowSerCOT}
\end{equation}
Here, $B_m$ with $m \in \mathbb{N}_0$ is a Bernoulli number
\cite[Section 23]{AbSt1972}.

If we now insert the leading terms of
(\ref{PowSerCOTH}) and (\ref{PowSerCOT}) into the
product $\coth (az) \cot (bz)$ and use $B_0 = 1$ and $B_2 = 1/6$
\cite[Eq.\ (23.1.3)]{AbSt1972}, we obtain:
\begin{equation}
\coth (az) \cot (bz) \; = \;
\frac{1}{abz^2} \, + \, \frac{a}{3b} \, - \, \frac{b}{3a}
\, + \, O \left( z^2 \right) \, , \qquad z \to 0 \, .
\end{equation}
This suggests -- consistent with the renormalisation 
of the effective Lagrangian -- the definition of the function
\begin{equation}
f (z) \; = \; a b \coth (az) \cot (bz) \, - \, \frac{1}{z^2}
\, - \, \frac{a^2 - b^2}{3} \, ,
\qquad a, b \in \mathbb{R} \, ,
\label{def_g}
\end{equation}
which corresponds to the left-hand side
of Eq.~(\ref{id2}) and which satisfies
\begin{equation}
f (z) \; = \; O \left( z^2 \right) \, , \qquad z \to 0 \, .
\label{asy_g_0}
\end{equation}

For the determination of the poles of $f (z)$, we use the
above equations (\ref{cotexpansion}) and (\ref{cothexpansion})
which can be reformulated as~\cite[Eq.\ (30:6.6)]{SpaOld1987}
\begin{equation}
\coth (z) \; = \; \sum_{m=-\infty}^{\infty} \,
\frac{z}{z^2 + m^2 \pi^2} \, ,
\qquad z \ne i k \pi \, , \quad k \in \mathbb{Z} \, ,
\label{ParFracCOTH}
\end{equation}
and \cite[Eq.\ (34:6.5)]{SpaOld1987}
\begin{equation}
\cot (z) \; = \; \sum_{m=-\infty}^{\infty} \,
\frac{z}{z^2 - m^2 \pi^2} \, ,
\qquad z \ne k \pi \, , \quad k \in \mathbb{Z} \, .
\label{ParFracCOT}
\end{equation}

From these expansions and from the definition of $f(z)$ we
conclude that $f (z)$ has the simple poles
\begin{eqnarray}
a_k & = & i \, \frac{k \pi}{a} \, , \qquad k \in \mathbb{Z}
\setminus \{ 0 \} \, ,
\label{PoleIm_g}
\\
a'_k & = & \phantom{i \,} \frac{k \pi}{b} \, , \qquad k \in \mathbb{Z}
\setminus \{ 0 \} \, .
\label{PoleRe_g}
\end{eqnarray}
Next, we want to determine the corresponding residues
\begin{equation}
b_k \; = \; \Res{z = a_k} f(z)
\; = \; \lim_{z \to a_k} \, \left[ (z-a_k) \, f (z) \right]
\label{def_ResIm_g}
\end{equation}
and
\begin{equation}
b'_k \; = \; \Res{z = a'_k} f(z)
\; = \; \lim_{z \to a'_k} \, \left[ (z-a'_k) \, f (z) \right] \, .
\label{def_ResRe_g}
\end{equation}
 
For that purpose, we rewrite (\ref{ParFracCOTH}) and (\ref{ParFracCOT})
as follows by isolating the terms that contribute to the residues at
$a_k$ and $a'_k$, respectively:
\begin{eqnarray}
\lefteqn{\coth (a z) \; = \;
\sum_{n=-\infty}^{\infty} \, \frac{z/a}{(z+i n\pi/a)(z-i n\pi/a)}}
\nonumber \\
& = & \sum_{\substack{n=-\infty \\ n \ne \pm k}}^{\infty} \,
\frac{z/a}{(z+i n\pi/a)(z-i n\pi/a)} \, + \,
\frac{2z/a}{(z+i k\pi/a)(z-i k\pi/a)} \, ,
\label{ParFrac_COTH}
\\
\lefteqn{\cot (b z) \; = \;
\sum_{m=-\infty}^{\infty} \, \frac{z/b}{(z+m\pi/b)(z-m\pi/b)}}
\nonumber \\
& = & \sum_{\substack{m=-\infty \\ m \ne \pm k}}^{\infty} \,
\frac{z/b}{(z+m\pi/b)(z-m\pi/b)} \, + \,
\frac{2z/b}{(z+k\pi/b)(z-k\pi/b)} \, .
\label{ParFrac_COT}
\end{eqnarray}
With the help of (\ref{ParFrac_COTH}), we then obtain:
\begin{eqnarray}
b_k & = & \lim_{z \to a_k} \, \left[ (z-a_k) \, f (z) \right]
\nonumber \\
& = &
\lim_{z \to ik\pi/a} \, \left[ (z-ik\pi/a) \,
 a b \coth (a z) \cot (b z) \right]
\nonumber \\
& = & \lim_{z \to ik\pi/a} \, \frac{2 b z \cot (b z)}{(z + i k \pi / a)}
\; = \; b \cot (i k \pi b / a) \, .
\end{eqnarray}
If we now use $\cot (i z) \; = \; - i \coth (z)$,
we finally obtain
\begin{equation}
b_k \; = \; - i b \coth (k \pi b / a) \, ,
\qquad k \in \mathbb{Z} \setminus \{ 0 \} \, .
\label{ResIm_g}
\end{equation}
Similarly, we obtain with the help of (\ref{ParFrac_COT}):
\begin{equation}
b'_k = \lim_{z \to a'_k} \left[ (z-a'_k) \, f (z) \right]
= a \cot (k \pi a / b) \, , \qquad k \in \mathbb{Z} \setminus \{ 0 \} .
\label{ResRe_g}
\end{equation}

If (\ref{f_MitLef}) is to be used for the
derivation of a partial-fraction decomposition for $f (z)$ defined by
(\ref{def_g}), it is natural to identify the closed
contour $\mathcal{C}_n$ with the rectangle having the 4 sides
$\mathcal{X}_{n}^{(\pm)}$ and $\mathcal{Y}_{n}^{(\pm)}$, where
\begin{equation}
\mathcal{X}_{n}^{(\pm)} \; = \; \bigl\{ z_n = x_n + iy_n
\big\vert x_n = (2s-1) X_n, 0 \le s \le 1, y_n = \pm Y_n \bigr\}
\label{SideX}
\end{equation}
and
\begin{equation}
\mathcal{Y}_{n}^{(\pm)} \; = \; \bigl\{ z_n = x_n + iy_n
\big\vert x_n = \pm
X_n, y_n = (2t-1) Y_n, 0 \le t \le 1 \bigr\} \, .
\label{SideY}
\end{equation}
Clearly, $X_n$ and $Y_n$ have to be chosen in such a way that the poles
$a_k = k \pi/b$ and $a'_k = i k \pi/a$ of $f (z)$ do not lie on this
rectangle.

We are on the safe side if we choose $X_n$ and $Y_n$ in such a way that
they are located in the middle between two neighbouring poles of $f (z)$:
\begin{equation}
X_n \; = \; \frac{\pi}{b} (n+1/2) \, , \qquad
Y_n \; = \; \frac{\pi}{a} (n+1/2) \, , \qquad n \in \mathbb{N} \, .
\end{equation}
We now have to show that $f (z)$ is bounded on the rectangle
$\mathcal{C}_n$ according to (iii) of Theorem 1. For that
purpose, we now use
\cite[Eq.\ (30:11.2)]{SpaOld1987}
\begin{equation}
\coth (x+iy) \; = \;
\frac{\sinh (2x) - i \sin (2y)}{\cosh (2x) - \cos (2y)}
\end{equation}
and \cite[Eq.\ (34:11.2)]{SpaOld1987}
\begin{equation}
\cot (x+iy) \; = \;
\frac{\sin (2x) - i \sinh (2y)}{\cos (2x) - \cosh (2y)} \, .
\end{equation}
If $s$ in (\ref{SideX}) satisfies $s = 1/2$, which implies $x_n = 0$ and
$y_n = \pm Y_n$, or $t$ in (\ref{SideY}) satisfies $t = 1/2$, which
implies $x_n = \pm X_n$ and $y_n = 0$, we obtain
\begin{equation}
\coth (\pm i a Y_n) \cot (\pm i b Y_n) \; = \;
\frac{- i \sin (\pm 2 a Y_n)}{[1 - \cos (\pm 2 a Y_n)]} \,
\frac{i \sinh (\pm 2 b Y_n)}{[1 - \cosh (\pm 2 b Y_n)]} \, ,
\end{equation}
which remains bounded on $\mathcal{C}_n$ as $n \to \infty$, or
\begin{equation}
\coth (\pm a X_n) \cot (\pm b X_n) \; = \;
\frac{\sinh (\pm 2 a X_n)}{[\cosh (\pm 2 a X_n) - 1]} \,
\frac{\sin (\pm 2 b X_n))}{[1 - \cos (\pm 2 b X_n)]} \, ,
\end{equation}
which also remains bounded on $\mathcal{C}_n$ as $n \to \infty$.

If $s$ in (\ref{SideX}) and $t$ in (\ref{SideY}) satisfy $s, t \ne 1/2$,
$f (z)$ remains bounded on the rectangles $\mathcal{C}_n$ because of the
periodicity and boundedness of $\cos$ and $\sin$ for real arguments, and
we find
\begin{equation}
\lim_{n \to \infty} \, \coth (a z_n) \; = \; \lim_{n \to \infty} \,
\frac{\sinh (2 a x_n)}{\cosh (2 a x_n)}
\end{equation}
and
\begin{equation}
\lim_{n \to \infty} \, \cot (b z_n) \; = \; i \, \lim_{n \to \infty} \,
\frac{\sinh (2 b y_n)}{\cosh (2 b y_n)} \, .
\end{equation}

Thus, the constant $M$ in (iii) of Theorem 1 can in the case of
$f (z)$ be chosen according to
\begin{equation}
M \; = \; \max_{n \in \mathbb{N}} \,
\bigl( \vert f (z_n) \vert \bigr) \, ,
\qquad z_n \in \mathcal{C}_n \, .
\end{equation}
Then, the $R_n$ satisfying (i) of Theorem 1 should be chosen
according to
\begin{equation}
R_n \; = \; \min (X_n, Y_n) \, ,
\end{equation}
and the constant $S$ in (ii) of Theorem 1 should be chosen
according to
\begin{equation}
S \; = \; 4 \max (b/a, a/b) \, .
\end{equation}

However, the function $f (z)$ defined in (\ref{def_g}) does not exactly
meet the requirements of Theorem 1. Instead of a sequence of
simple poles $a_1$, $a_2$, $a_3$, \ldots with corresponding residues
$b_1$, $b_2$, $b_3$, \ldots, there are actually two sequences $a_{\pm
1}$, $a_{\pm 2}$, $a_{\pm 3}$, \ldots of simple poles with corresponding
residues $b_{\pm 1}$, $b_{\pm 2}$, $b_{\pm 3}$, \ldots, which implies
that we have a partial-fraction decomposition of the following kind:
\begin{equation}
F (z) \; = \; F (0) \, + \, \sum_{m=1}^{\infty} \,
\left\{ \frac{b_m}{z - a_m} + \frac{b_m}{a_m} \right\} \, + \,
\sum_{m=1}^{\infty} \,
\left\{ \frac{b_{-m}}{z - a_{-m}} + \frac{b_{-m}}{a_{-m}} \right\} \, .
\label{ModMittLef_1}
\end{equation}
If we take into account that the hyperbolic cotangent is \emph{odd}
according to $\coth (-z) = - \coth (z)$, we see from (\ref{PoleIm_g}) and
(\ref{ResIm_g}) or from (\ref{PoleRe_g}) and (\ref{ResRe_g}) that the
poles and residues of $F (z)$ have to satisfy for all $k \in \mathbb{N}$
the (anti-)symmetry relations
\begin{equation}
a_k \; = \; - \, a_{-k} \, , \qquad b_k \; = \; - \, b_{-k} \, .
\end{equation}

If we also take into account that (\ref{asy_g_0}) implies
\begin{equation}
f (0) \; = \; 0 \, ,
\end{equation}
we see that (\ref{ModMittLef_1}) is to be replaced by the following
partial-fraction decomposition:
\begin{equation}
F (z) \; = \; \sum_{m=1}^{\infty} \,
\left\{ \frac{b_m}{z - a_m} \, - \, \frac{b_m}{z + a_m} \, + \,
\frac{2 b_m}{a_m} \right\} \, .
\label{ModMittLef_2}
\end{equation}

If we now insert the poles (\ref{PoleIm_g}) and (\ref{PoleRe_g}) and the
corresponding residues (\ref{ResIm_g}) and (\ref{ResRe_g}),
respectively, into (\ref{ModMittLef_2}), we obtain the following
partial-fraction decomposition:
\begin{eqnarray}
\lefteqn{ab \coth (az) \cot (bz) \, - \, \frac{1}{z^2} \, - \,
\frac{a^2 - b^2}{3}} \nonumber \\
& = &  \sum_{k=1}^{\infty} \, a \coth (k \pi a / b) \,
\left\{
\frac{1}{z - k\pi/b} - \frac{1}{z + k\pi/b} + \frac{2}{k\pi/b}
\right\} \nonumber \\
& & - \, \sum_{k=1}^{\infty} \, i b \coth (k \pi b / a) \,
\left\{
\frac{1}{z - ik\pi/a} - \frac{1}{z + ik\pi/a} + \frac{2}{ik\pi/a}
\right\} \, .
\label{g_exp_1}
\end{eqnarray}
By putting the expressions in curly brackets on a common denominator, we
obtain (\ref{id2}).

%
%
\section{Electric--Magnetic Duality}
\label{EMDuality}
Electric-magnetic duality, understood as a mutual transformation of
electric and magnetic quantities, has attracted a great deal of attention
since Dirac's ideas on magnetic monopoles~\cite{Co1982,Bl1988} were
introduced in classical electrodynamics. 
Whereas electric-magnetic duality can be formulated as a
continuous symmetry of the classical Maxwell equations~\cite{Jackson}
(either including both types of charges or without any charges), we
devote this section to a brief study of discrete duality
transformations of quantum effective actions. In particular, we
consider two types of duality:
\begin{eqnarray} 
\text{Type\,\,I:}&&\bm{E}\to\bm{B}, \quad \bm{B}\to
\bm{-E}, \nonumber\\
\text{Type\,\,II:}&&a\to-{\mathrm i} b,\quad b\to {\mathrm i} a. \label{dual1}
\end{eqnarray}
(In a Lorentz frame where $\bm{E}\,\| \bm{B}$, the Type II
duality implies $|\bm{B}|\to -{\mathrm i} |\bm{E}|$ and $|\bm{E}|\to
{\mathrm i} |\bm{B}|$.) As a consequence, the invariants transform as
\begin{eqnarray}
\text{Type\,\,I:}&&{\cal F}\to-{\cal F}, \quad {\cal G}\to -{\cal G},
\quad\Longrightarrow\quad b\to a, \quad a\to -b   \nonumber\\ 
\text{Type\,\,II:}&&{\cal F}\to{\cal F}, \quad {\cal G}\to{\cal
  G}. \label{dual2} 
\end{eqnarray}
Note that the duality of Type II preserves the invariants, so that
Maxwell's equations derivable from ${\cal L}_{\text{cl}}=-{\cal F}$ are
trivially invariant; by contrast, the classical Lagrangian is not
invariant under Type I, but Maxwell's equations are.\footnote{The
  transformations of $a$ and $b$ under Type I follow from
  Eqs.~(\ref{DefF},\ref{DefG}). Note that Eq.~(\ref{Defab}) is not
  meant to participate in the duality transformations, owing to the
  nonlinear relation between the two sets of invariants.} 
It therefore becomes obvious that only Type I affects physical
quantities (field strengths), whereas Type II signifies a certain
redundancy in the parameterisation of physical quantities.

In fact, this redundancy exists even for a larger class of duality
transformations of Type II,
\begin{equation}
a \to \pm {\mathrm i} \, b \, \quad
b \to \mp {\mathrm i}\, a. \label{dual3}
\end{equation}
which preserves the invariants ${\cal F}$ and ${\cal G}$. Since the
effective Lagrangian of QED is gauge and Lorentz invariant, the
effective Lagrangian for constant fields is necessarily a function of
${\cal F}$ and ${\cal G}$ only, ${\cal L}_{\text{eff}} = {\cal
L}_{\text{eff}} ({\cal F}, {\cal G})$, implying that the duality
transformations of Eq.~(\ref{dual3}) leave ${\cal L}_{\text{eff}} ({\cal
F}, {\cal G})$ invariant. This statement is not at all tied to
perturbation theory, and the duality (\ref{dual3}) holds therefore to
all orders in the external fields and to all loop orders. 
From a diagrammatic perspective, each external leg of a
diagram contributing to any given loop order corresponds to a field
strength tensor; and any possible contraction can be expressed in terms
of ${\cal F}$ and ${\cal G}$, ensuring this duality invariance. In some
sense, these invariants are therefore more fundamental than $a$ and $b$.

Let us now study the invariance properties of the effective Lagrangian
under consideration. 
From Eq.~(\ref{EffAct}), it is obvious that $\Delta{\cal L}$ is not 
invariant under Type I, similarly to the Maxwell
Lagrangian, (even a rotation of the contour in the first
quadrant does not help): in an asymptotic expansion, terms of odd
order in ${\cal F}$, ${\cal G}$ flip sign.

At a first glance, $\Delta{\cal L}$ indeed seems to be invariant under
the transformation (\ref{dual3}) as expected, since the integrand is
invariant.\footnote{The integrand is even invariant under a larger
class of duality transformations, including $a \to \pm {\mathrm i} b$,
$b \to \pm {\mathrm i}\, a$, owing to parity invariance of QED.}
Nevertheless, it has recently been argued~\cite{ChPa2001,ChPa2001prd}
that $\Delta{\cal L}$ is uniquely invariant only under Type II,
excluding explicitly the remaining transformations of
Eq.~(\ref{dual3}). The argument is based on the special function
representation (\ref{UniSpecial}) in terms of incomplete Gamma
functions.  

In our language, the argument could be rephrased in the following way:
Taking the details of the integral contour specified by $\eta$ and
$\epsilon$ into account corresponds to a modification $a \to
a\,\exp({\mathrm i}\,\delta)$, where $\delta$ is a small positive
quantity. We discuss the apparent uniqueness of the replacement $a \to
-{\mathrm i}\,b$ by way of example (see also Appendix~\ref{modelex}).

When applying the replacement $a \to a\,\exp({\mathrm i}\,\delta)$ to
$b_n$ in Eq.~(\ref{UniSpecial}),
the relevant factor in the argument of the last Gamma function 
in $b_n$ assumes the form ${\mathrm i}/{a} \to
({{\mathrm i}}/{a}) \exp(-{\mathrm i}\,\delta)$,
which has complex argument $\pi/2 - \delta$.
With attention to the fact that the incomplete Gamma
function has singularities along the negative real axis,
it therefore becomes apparent that this argument could only 
be increased by an amount of $+\pi/2$ (not $-\pi/2$), if we want
to stay on the same branch of the incomplete Gamma function. 
This seems to fix uniquely the replacement $a \to - {\mathrm i}\, b$
because in this case,
\begin{equation}
\frac{{\mathrm i}}{a}\,\exp(-{\mathrm i}\,\delta) \to
\frac{{\mathrm i^2}}{b}\,\exp(-{\mathrm i}\,\delta) \to
-\frac{1}{b}\,\exp(-{\mathrm i}\,\delta) =
-\frac{1}{b} + {\mathrm i}\, \gamma,
\end{equation}
in agreement with the last term in the definition of $c_n$ in
Eq.~(\ref{UniSpecial}). Here, $\gamma$ denotes a further
infinitesimal positive quantity.

The contradiction between the required general form of the duality
(\ref{dual3}) and the seemingly unique form of Type II can be resolved
by looking at the integral representation: consider the representation
(\ref{EffAct}) of $\Delta{\cal L}$ as a function of 
two complex variables $z_1$ and $z_2$: 
\begin{equation}
\label{z1z2}
\Delta {\mathcal L}(z_1, z_2) = - \frac{e^2}{8 \pi^2} \,
\lim_{\epsilon,\eta \to 0^{+}} 
\int\limits_{{\mathrm i}\,\eta}^{\infty + {\mathrm i}\,\eta} 
\frac{{\rm d} s}{s} {\rm e}^{- (m^2 - {\rm i} \epsilon)\, s}\, 
\biggl[  \, z_1 z_2 \, \coth(e z_1 s) \, \cot(e z_2 s) \, -
\frac{z_1^2 - z_2^2}{3} - \frac{1}{(e s)^2} \biggr] \,.
\end{equation}
The duality of Type II (\ref{dual1}) can be formulated as 
the identity $\Delta {\mathcal L}(a,b) = \Delta {\mathcal L}(-{\mathrm
i}\,b, {\mathrm i}\, a)$. Because $a$ and $b$ are real and positive,
this identity is valid in a strict sense only if the function $\Delta
{\mathcal L}(z_1, z_2)$ is one-valued in the relevant range of the 
complex arguments of $z_1$ and $z_2$.

In order to avoid the poles given by the 
cotangent and hyperbolic cotangent functions, 
the complex arguments of $z_1$
and $z_2$ in (\ref{z1z2}) cannot be varied without restriction if we
keep the class of allowed contours fixed (recall that the limits of
integration in (\ref{EffAct}) can also be chosen as ${\mathrm
i}\,\eta$ and $\infty + {\mathrm i}\,\eta$, respectively).
Taking into account the infinitesimal parts $\epsilon$
and $\eta$, we are led to conclude that the 
integral representation remains valid for the fixed class of contours
in the argument range
\begin{eqnarray}
\label{restriction}
-\pi/2 & \leq & \arg (z_1) < \pi/2\, , \nonumber\\[2ex]
0 & \leq & \arg (z_2) < \pi\, 
\end{eqnarray}
for $z_1$ and $z_2$ (observe the fine difference between the $<$ and
$\leq$ signs!). The restriction given in (\ref{restriction})
identifies (as a function of $z_1$ and $z_2$) a ``physical'' or
``causal'' branch of the effective action for a given contour.
Among the four different replacements
$\Delta {\mathcal L}(a,b) \to 
\Delta {\mathcal L}(\pm {\mathrm i}\,b, \pm {\mathrm i} a)$
or $\Delta {\mathcal L}(a,b) \to 
\Delta {\mathcal L}(\pm {\mathrm i}\,b, \mp {\mathrm i} a)$,
it is only the replacement 
$\Delta {\mathcal L}(a,b) \to
\Delta {\mathcal L}(-{\mathrm i}\, b, {\mathrm i}\,a)$
which respects the restriction set
by (\ref{restriction}) for a fixed contour. Therefore, the duality of
Type II is only ``unique'' in connection with a precisely specified
integral representation which does not allow for an unrestricted
analytic continuation of its arguments. In other words, the seeming
``uniqueness'' is simply a shortcoming of the integral representation
(and also of the special-function representation being identical to
the former).

In order to achieve full invariance under the dualities (\ref{dual3}),
we have to allow for the fact that the contour also has to be
readjusted (or certain poles are allowed to be crossed without picking
up their contribution). From a different perspective, it is only
natural to perform the duality transformation first, and then specify
the details of the contour. This is perfectly justified, since the
particular choice of the contour is not a {\em result} of the
calculation (of the fermion determinant in this case), but rather an
additional piece of information that has to be inserted afterwards in
order to define the integral. This information arises, of course, from
physical requirements: the $\pm {\mathrm i} \epsilon$ prescription is dictated
by causality, and the shift by $\eta$ can, e.g., be fixed by requiring
that the pair-production probability related to the imaginary part of
$\Delta{\cal L}$ is a number between 0 and 1. The resulting integral
representation will always lead to the same special-function
representation (\ref{UniSpecial}). 

From another point of view, the special-function
representation and the integral representation with fixed contour
remove a part of the above-mentioned redundancy in the parameterisation
of the field strength invariants, which is generally present in the
effective Lagrangian. 

At this point, let us remark that dualities of Type II or
(\ref{dual3}) can be very useful from a technical viewpoint, although
they have no physical meaning: for instance, from the effective
Lagrangian for a purely magnetic field, one can extract information
about the electric case with the aid of the substitution,
$|\bm{B}|\to -{\mathrm i} |\bm{E}|$ \cite{DiTsZi1979}. Moreover,
standard-model calculations in constant electromagnetic fields can
always be checked by testing their dual invariance; e.g. in
\cite{GiSh2000,GiSc2001}, this dual invariance is visible in neutrino
amplitudes in electromagnetic fields. Of course, to be on the safe
side, the duality of Type II suffices for such a check in order to
avoid problems of the kind mentioned before.

The question of electric-magnetic duality becomes even more
interesting for systems which are characterised by additional Lorentz
covariant quantities. For example, let us consider QED with constant
fields in a heat bath; the latter involves an additional Lorentz
vector $u_\mu$, the heat-bath four-velocity, allowing for one further
gauge and Lorentz invariant quantity [cf. Eq.~(3.143)
of~\cite{DiGi2000}]:   
\begin{equation}
{\mathcal E} = (u_\mu F^{\mu\nu})^2 = 
u_\mu F^{\mu\nu} u^\rho F_{\rho\nu}. \label{calE}
\end{equation}
Under duality of Type II, ${\mathcal E}$ is trivially invariant, since
it is linearly independent of ${\mathcal F}$ and ${\mathcal G}$, and
thereby independent of $a$ and $b$. In fact, the known QED effective
actions, depending on $a$, $b$ and ${\mathcal E}$ at finite temperature,
show this invariance under Type II~\cite{DiGi2000}.  

Under duality of Type I, ${\mathcal E}$ transforms according to
${\mathcal E} \to {\mathcal E} +2 {\mathcal F} $; and similarly to the
zero-temperature case, the finite-temperature effective action is
generally not invariant under Type I. Nevertheless, the dominant
low-temperature contribution arising at two-loop order astonishingly
exhibits an invariance under Type I. This might be related to the fact
that only transversal thermal fluctuations of the photon give rise to
this contribution.

%
%
\section{Acceleration of Convergence}
\label{AccConv}

From the asymptotic expansion of the incomplete Gamma
function as $z \to \infty$~\cite[Eq. (6.5.32)]{AbSt1972},
\begin{equation}
\label{GammaAsymp}
\mathrm{e}^z \, \Gamma(a, z) = z^{a-1} \, 
\left[ 1 + \frac{a-1}{z} + \frac{(a-1)(a-2)}{z^2} + 
{\mathcal O}(z^{-3}) \right]\, , 
\quad \vert \arg z \vert < 3\pi/2 \, ,
\end{equation}
we obtain
\begin{equation}
\frac{\mathrm{e}^{n z} \Gamma(0, n z)}{n} \; = \; 
\frac{1}{n^2 z} - \frac{1}{n^3 z^2} + \mathrm{O} \left( n^{-4} \right) 
\, , \qquad n \to \infty \, .
\end{equation}
Since the cotangents in ({\ref{UniSpecial}) rapidly approach one as $n
\to \infty$, we can conclude that terms in ({\ref{UniSpecial}) are 
essentially of order $\mathrm{O} (n^{-2})$ as $n \to \infty$. A more
detailed analysis of the large-order behaviour of the terms in
({\ref{UniSpecial}) can be found in Appendices A and B of
\cite{VaLaMi1993}.

An $\mathrm{O} (n^{-2})$ behaviour of the terms also occurs in the
Dirichlet series $\zeta (s) = \sum_{n=1}^{\infty} n^{-s}$ for the
Riemann zeta function with $s = 2$. This is a very discomforting
observation since the Dirichlet series for $\zeta (2)$ converges quite
slowly -- it can be shown that increasing the number of terms of its
partial sum by a factor of 10 improves the accuracy by a single decimal
digit only -- and in the literature on convergence acceleration it is
one of the standard test systems for checking the capability of a
transformation in the case of a slowly convergent monotonic series.

Thus, the series expansions (\ref{RepSpecReal}) and
({\ref{UniSpecial}) represent typical cases of monotonic series which
converge so slowly that a straightforward evaluation by adding up one
term after another is computationally very unattractive. Of course,
the acceleration of the convergence of series of that kind has been
studied quite extensively in the literature and many techniques are
known to improve the efficiency of numerical computations (see for
example~\cite{We1989,BreRZa1991,We2001} and references therein).

Nevertheless, one should not forget that the acceleration of the
convergence of a slowly convergent monotonic series can be a very
challenging problem. Moreover, numerical instabilities due to rounding
errors are likely to occur. A sequence transformation accelerates
convergence by extracting and utilising information from a finite set of
input data on the index-dependence of the truncation errors. This is
normally accomplished by forming higher weighted differences. If the
input data are the partial sums of a \emph{strictly alternating} series,
the formation of higher weighted differences is a remarkably stable
process, but if the input data all have the \emph{same sign}, numerical
instabilities are quite likely. Thus, if the sequence to be transformed
are the partial sums of a slowly convergent monotonic series, numerical
instabilities are to be expected, and most convergence acceleration
methods are not able to obtain transformation results that are close to
machine accuracy.

In some cases, these instability problems can be overcome with the
help of a condensation transformation attributable to Van Wijngaarden,
which converts input data having the same sign to the partial sums of
an alternating series, whose convergence can be accelerated more
effectively (compare, for instance, Appendix A
of~\cite{JeMoSoWe1999}).  The condensation transformation was first
mentioned in \cite[pp.\ 126 - 127]{NaPhyLa1961} and only later
published by Van Wijngaarden \cite{vW1965}. It was used by Daniel
\cite{Da1969} in combination with the Euler transformation, and
recently, it was rederived by Pelzl and King \cite{PeKi1998}. Since
the transformation of a strictly alternating series by means of
nonlinear sequence transformations is a stable process, it was in this
way possible to evaluate special functions that are defined by
extremely slowly convergent monotonic series, not only relatively
efficiently but also close to machine accuracy \cite{JeMoSoWe1999}, or
to perform extensive quantum electrodynamical
calculations~\cite{JeMoSo1999,JeMoSo2001}.  Unfortunately, the use of
this ``combined nonlinear-condensation transformation''
(CNCT)~\cite{JeMoSoWe1999} is not always possible: The conversion of a
monotonic to an alternating series requires that terms of the input
series with large indices can be computed.

However, this CNCT is well suited for the acceleration of the
convergence of the series expansions (\ref{RepSpecReal}) and
({\ref{UniSpecial}) considered in this article.

\begin{table}[htb]
\begin{center}
\begin{minipage}{14cm}
\begin{center}
\caption{Evaluation of the real (dispersive) part of the 
QED effective Lagrangian for 
$|\bm{E}| = 30\, E_{\mathrm cr}$
and $|\bm{B}| = 30\, B_{\mathrm cr}$
by evaluating the 
special function representation (\ref{RepSpecReal}) using 
a combination~\cite{JeMoSoWe1999} of the Van Wijngaarden condensation 
transformation defined in Eqs.~(\ref{VWSerTran}) --  (\ref{B2a})
and the nonlinear delta transformation~\cite{We1989}.
The ${\bf S}_{n}$ are defined in (\ref{PsumS}), and the nonlinear
delta transform ${\delta}_{n}^{(0)} \bigl(1, {\bf S}_0 \bigr)$
is defined in Eq.~(8.4-4) of~\cite{We1989}.
The result is given in units of $m^4$. The apparent convergence 
of the delta transforms in the second column is indicated by 
underlining. After 21 iterations, the transforms
${\delta}_{n}^{(0)} \bigl(1, {\bf S}_0 \bigr)$ have stabilised 
to the 15-figure result $8.393~398~582~100~617$.}
\label{example}
\begin{tabular}{crr}%
\hline
\hline
$n$%
& \multicolumn{1}{c}{${\bf S}_{n}$}%
& \multicolumn{1}{c}{${\delta}_{n}^{(0)} \bigl(1, {\bf S}_0 \bigr)$}%
\\
\hline
0 &   11.834~710~587~368 &  11.834~710~587~368 \\
1 &    6.388~353~476~842 &  \underline{8}.463~873~830~587 \\
2 &    9.741~830~922~440 &  \underline{8.3}84~963~963~657 \\
3 &    7.413~293~009~436 &  \underline{8.393}~553~703~382 \\
4 &    9.141~317~648~944 &  \underline{8.393~3}98~289~155 \\
5 &    7.803~272~984~326 &  \underline{8.393~39}9~592~701 \\
6 &    8.870~519~831~631 &  \underline{8.393~398}~337~299 \\
7 &    8.000~385~721~330 &  \underline{8.393~398}~666~561 \\
8 &    8.721~936~912~365 &  \underline{8.393~398~5}94~227 \\
9 &    8.115~447~030~287 &  \underline{8.393~398~58}3~148 \\
10 &    8.630~906~154~429 & \underline{8.393~398~58}0~473 \\
11 &    8.188~731~860~505 & \underline{8.393~398~582}~040 \\
12 &    8.571~048~192~682 & \underline{8.393~398~582}~143 \\
13 &    8.238~223~421~331 & \underline{8.393~398~582~1}04 \\
14 &    8.529~698~152~551 & \underline{8.393~398~582~09}7 \\
15 &    8.273~084~942~867 & \underline{8.393~398~582~099} \\
16 &    8.500~073~809~529 & \underline{8.393~398~582~099} \\
\hline
\hline
\end{tabular}
\end{center}
\end{minipage}
\end{center}
\end{table}

The method consists in first rewriting the slowly convergent monotonic
input series $\sum_{k=0}^{\infty} \tau (k)$ into an alternating
series. In the second step, the convergence of the alternating series is
accelerated via a suitable nonlinear sequence transformation.  In our
case, we have $\tau (k) = [a_{k+1} + d_{k+1}]$, with $a_{k+1}$ and
$d_{k+1}$ given by Eq.~(\ref{RepSpecReal}). The partial sums
\begin{equation}
\sigma_n \; = \; \sum_{k=0}^{n} \, \tau (k)
\label{ParSumInput}
\end{equation}
increase monotonically, i.e.~$\sigma_{n+1} > \sigma_n$ for all
$n=0,\dots,\infty$ if all terms satisfy $\tau (n) \ge 0$. Let us further
assume that the sequence of the partial sums
$\{\sigma_n\}_{n=0}^{\infty}$ converges to some limit $\sigma =
\sigma_{\infty}$ as $n \to
\infty$. Following Van Wijngaarden \cite{vW1965}, we transform the original
series into an alternating series $\sum_{j=0}^{\infty} (-1)^j {\bf
A}_j$ according to
\begin{eqnarray}
\sum_{k=0}^{\infty} \, \tau (k) & = &
\sum_{j=0}^{\infty} \, (-1)^j \, {\bf A}_j \, ,
\label{VWSerTran}
\\
{\bf A}_j & = & \sum_{k=0}^{\infty} \, {\bf b}_{k}^{(j)} \, ,
\label{A2B}
\\
{\bf b}_{k}^{(j)} & = & 2^k \, \tau (2^k\,(j+1)-1) \, .
\label{B2a}
\end{eqnarray}
The terms ${\bf A}_j$ defined in Eq.\ (\ref{A2B}) are all positive if
the terms $\tau (k)$ of the original series are all positive. The
quantities ${\bf A}_j$ are commonly referred to as the condensed sums,
and the series $\sum_{j=0}^{\infty} (-1)^j {\bf A}_j$ is referred to as
the Van Wijngaarden transformed series.

The transformation from a monotonic series to a strictly alternating
series according to Eqs.~(\ref{VWSerTran}), (\ref{A2B}) and
(\ref{B2a}) is essentially
a reordering of the terms $\tau (k)$ of the original series. 
We define the partial sums
\begin{equation}
\label{PsumS}
{\bf S}_n \; = \; \sum_{j=0}^{n} \, (-1)^j \, {\bf A}_j
\end{equation}
of the Van Wijngaarden transformed series. As 
illustrated in Table~1 of~\cite{JeMoSoWe1999},
the ${\bf S}_n$ with $n \ge 0$ reproduces the partial sum $\sigma_n$,
Eq. (\ref{ParSumInput}), which contains the first $n+1$ terms of the
original series. 
Formal proofs of the correctness of this rearrangement
can be found in Ref.\ \cite{Da1969} or in the Appendix of Ref.\
\cite{PeKi1998}.

The series (\ref{A2B}) for the terms of the Van Wijngaarden transformed
series can be rewritten as follows:
\begin{equation}
{\bf A}_j \; = \; \tau (j) \, + \, 2 \tau (2j+1) \, + \, 
4 \tau (4j+3) \, + \,
\ldots \, .
\end{equation}
Since the terms $\tau (k)$ of the original series have by assumption the
same sign, we immediately observe
\begin{equation}
\vert {\bf A}_j \vert \ge \vert \tau (j) \vert \, .
\label{aBoundA}
\end{equation}
Consequently, the Van Wijngaarden transformation, given by
Eqs.~(\ref{VWSerTran}), (\ref{A2B}) and (\ref{B2a}), does not lead to an
alternating series whose terms decay more rapidly in magnitude than the
terms of the original monotonic series. However, an acceleration of
convergence may be achieved if the partial sums (\ref{PsumS}) of the Van
Wijngaarden transformed series are used as input data in a convergence
acceleration process, and -- as, for example, discussed in Appendix A
of~\cite{JeMoSoWe1999}) -- the convergence of alternating series can be
accelerated much more effectively than the convergence of monotonic
series.  For the acceleration of convergence we use the delta
transformation given in Eq.~(8.4-4) of~\cite{We1989}, which was found to
be a very effective accelerator for Van Wijngaarden transformed
series~\cite{JeMoSoWe1999}.

This will now be demonstrated explicitly.
From a consideration of the expressions
(\ref{RepSpecReal}) and ({\ref{UniSpecial}) it is obvious that
the computationally most demanding special cases will be encountered
for large fields; in these cases, many terms of the 
representation (\ref{RepSpecReal}) are needed in order
to achieve convergence [accordingly, for 
strong fields we encounter problematic oscillations 
in the integrand of Eq.~(\ref{EffAct})].
We consider only one example here -- Table~\ref{example} --,  
which describes the evaluation of 
the dispersive (real) part of the effective Lagrangian 
at field strength $|\bm{E}| = b = 30\, E_{\mathrm cr}$
and $|\bm{B}| = a = 30\, B_{\mathrm cr}$.
This does not preclude the possibility that other efficient
calculational methods exist for the evaluation of~(\ref{RepSpecReal}).
However, we stress here that a suitable acceleration of the 
convergence of the special function representations 
(\ref{RepSpecReal}) and (\ref{UniSpecial}) removes the 
principal numerical difficulty associated with the 
slow convergence of the series at large field strength.
In our example -- see Table~\ref{example} --,
the highest index encountered in the calculation is
$\hat{n} = 37748736$, the total number of evaluations
of terms $[a_n + d_n]$ is $405$. The ratio of these two
numbers is roughly $93000$, which corresponds to an acceleration
of the calculation by roughly five orders of magnitude. 

The CNCT transforms the slowly convergent series
expansions (\ref{RepSpecReal}) and
({\ref{UniSpecial}) into the rapidly converging sequence
of the delta transforms ${\delta}_{n}^{(0)} \bigl(1, {\bf S}_0 \bigr)$
(see Table~\ref{example}). In general, no direct
interpretation is available for the delta transforms~\cite{We1989},
just as much as Pad\'{e} approximants~\cite{Wy1956b} lack a direct physical 
interpretation. At best, the delta transforms can be viewed
as the analytic continuations (``extrapolations'') of the partial  
sums (\ref{PsumS}) of the Van Wijngaarden transformed alternating
series ${\bf S}_n$ to $n \to \infty$.

%
%
\section{Conclusion}

We have investigated questions related to the representation 
of the quantum electrodynamic (QED) effective Lagrangian 
and its numerical evaluation.
In Sec.~\ref{RepSpecial}, we recall our previous 
results given in~\cite{VaLaMi1993,VaMiLa1994} 
for special function representations of the
effective Lagrangian, and we clarify the mathematical
notation used in the special function representations
(\ref{RepSpecReal}), (\ref{RepSpecIm}) and (\ref{UniSpecial}).
The representation (\ref{UniSpecial}) unifies 
the real and imaginary parts.

In Sec.~\ref{ImpIdentity}, we present a detailed 
description of the proof of a certain mathematical
identity (\ref{id}) 
used in our investigations~\cite{VaLaMi1993,VaMiLa1994}.
In Sec.~\ref{EMDuality}, we discuss the question as to whether the
QED effective Lagrangian is invariant under certain types of
electric-magnetic duality. We conclude that gauge and Lorentz
symmetry guarantees invariance under a general class of dualities
(\ref{dual3}) to all loop orders; but discrete representations of the
effective Lagrangian may not realize this general dual invariance in a
strict sense, so that only a smaller duality subgroup (Type II) remains.

In Sec.~\ref{AccConv}, we show that the convergence over the summation 
index $n$ of the special function representation (\ref{RepSpecReal})
can be accelerated efficiently by the application of the
CNCT transformation~\cite{JeMoSoWe1999}. In this way, the computing time is 
reduced by several orders of magnitude. Based on the results of the 
current paper, we expect to carry out detailed studies related
to various projected and ongoing experiments and astrophysical 
phenomena~\cite{DiGi2000,St1979,VaBa1980,DuTh1992,Ba1995} involving
strong static-field conditions (or fields with frequencies that
are small as compared to the electron Compton wavelength). 

%
%
\section*{Acknowledgements}

The authors wish to acknowledge very helpful conversations with 
Professors G.~Soff, W.~Dittrich, D.~G.~C.~McKeon, and V.~Elias.
H.G. acknowledges financial support by the Deutsche Forschungsgemeinschaft
under contract Gi 328/1-1. S.R.V. acknowledges a supporting grant from 
the Natural Sciences, Engineering and Research Council of Canada (NSERC).

\appendix

\section{Model Example}
\label{modelex}

Consider as a model example the integral
\begin{equation}
\label{ModelExample}
{\cal M}(a,b) =
\int_{{\mathrm i}\,\eta}^{\infty + {\mathrm i}\,\eta} 
{\mathrm d}t\, \exp(-t) \,
\left( \frac{1}{1 - {\mathrm i}\, a \,t} + 
\frac{1}{1 + {\mathrm i}\, a \,t} +
\frac{1}{1 - b \,t} + 
\frac{1}{1 + b \,t}\right),
\end{equation}
where $a$ and $b$ are {\em a priori} real variables, 
but can be also be generalised to the complex case
$a \to z_1$, $b \to z_2$.
The integrand is invariant under the four different
``duality transformations'' $a \to \pm {\mathrm i} b$,
$b \to \pm {\mathrm i} a $ ($\mp {\mathrm i} a$), but the integral is 
not. Owing to the poles of the integrand,
the imaginary parts change sign 
when the complex arguments $a$ and $b$ are varied such 
that one of the poles of the integrand is crossed.
When considered as a function of complex arguments
$z_1$ and $z_2$, the function
${\cal M}(z_1, z_2)$ has branch cuts along the positive and negative
imaginary $z_1$ axis and along the positive and negative real 
$z_2$ axis. Furthermore, by inspection of the integrand we conclude
that the argument ranges for $z_1$ and $z_2$ are
\begin{eqnarray}
\label{ModelRestrictions}
-\pi/2 & \leq & \arg (z_1) < \pi/2\,, \nonumber\\
0 & \leq & \arg (z_2) < \pi\,.
\end{eqnarray} 
From these relations it becomes clear that if we want to stay
on the principal branch in the complex $z_1,z_2$ plane, then 
we have to modify the arguments of $a$ and $b$ in accord with 
the restrictions given by (\ref{ModelRestrictions}). This singles
out the duality of Type II 
\begin{equation}
a \to - {\mathrm i}\,b\,, \quad b \to {\mathrm i}\,a\,.
\end{equation}
We conclude the discussion of the model example by pointing out 
that it can be expressed as 
\begin{eqnarray}
{\cal M}(a,b) &=& \lim_{\epsilon \to 0^{+}} \, \left\{
\frac{{\mathrm i}}{a}\,
\left[ \exp\left(\frac{{\mathrm i}}{a}\right)\,
\Gamma\left(0, \frac{{\mathrm i}}{a}\right) - 
\exp\left(- \frac{{\mathrm i}}{a}\right)\,
\Gamma\left(0, -\frac{{\mathrm i}}{a}\right) \right] \right.
\nonumber\\[1ex]
& & \;\;\; \left. + \frac{1}{b}\,
\left[ \exp\left(\frac{1}{b}\right)\,
\Gamma\left(0, \frac{1}{b}\right) -  
\exp\left(- \frac{1}{b}\right)\,
\Gamma\left(0, -\frac{1}{b} + {\mathrm i}\,\epsilon\right) \right]
\right\}.
\end{eqnarray}
There is a certain analogy to the special function representation
(\ref{UniSpecial}).

\end{document}